# ASTRONOMICAL CHEMICAL EVOLUTION CAPABILITY FROM GRAPHENE TO POLYCYCLIC AROMATIC HYDROCARBON REPRODUCING OBSERVED INFRARED SPECTRUM

NORIO OTA
Graduate School of Pure and Applied Sciences, University of Tsukuba,
1-1-1 Tenoudai Tsukuba-city 305-8571, Japan; n-otajitaku@nifty.com

Interstellar ubiquitous infrared spectrum (IR) due to polycyclic aromatic hydrocarbon (PAH) was observed in many astronomical dust clouds. In a previous quantum chemistry study, it was suggested that PAH's having carbon pentagon-hexagon coupled molecules show good coincidence with observed IR. In order to clarify such coincidence, a capable astronomical chemical evolution path from graphene to PAH was studied based on the first principles calculation. Step 1 is a nucleation of nano-carbon after super-nova expansion by the super-cooling at expanding helium sphere. As a typical model, graphene molecule $C_{24}$ having coronene skeleton with seven carbon hexagons was tried. Nucleated graphene would be ejected to space as an important candidate of dust. Step 2 is a proton $H^+$ sputtering and passivation on ejected graphene molecule. Slow speed proton with energy less than 4.3eV makes hydrogenation, Graphene molecule $C_{24}$ was transformed to PAH $C_{24}H_{12}$. Higher speed proton having sufficient energy larger than 18.3 eV could make a void in a molecule as like $C_{23}H_{12}$. Molecular structure was transformed rapidly to a stable state by a quantum dynamics. Resulted structure was a combination of two carbon pentagons and five hexagons. Step 3 is photo-ionization of those molecules by high energy photon. Electrons are removed to make a molecule to cation. There may be two cases as light source, one is supernova decayed light, another is new born star illumination. Model molecule $C_{23}H_{12}$ became $C_{23}H_{12}^{+}$, $C_{23}H_{12}^{2+}$ and so on. Typical energy difference between such cation was 6.5 and 10.8 eV. If the light source has a nature of black-body radiation, effective temperature will be 18000K ~ 24000K, which suggested that central light source star may have 4 to 7 times heavier than our sun. Finally, theoretical IR spectrum was obtained for such cation. Again, very nice coincidence with observed IR was obtained. Especially in case of $C_{23}H_{12}^{2+}$, calculated emission spectrum revealed that among 13 major peaks 11 one could correlate with ubiquitous observed IR peaks.

## 1, INTRODUCTION

It is well known that interstellar ubiquitous infrared spectrum (IR) due to polycyclic aromatic hydrocarbon (PAH) was observed in many astronomical dust clouds from 3 to 20μm (Boersma et al. 2013, 2014). In previous quantum chemical calculations (Ota 2014, 2015a, 2015b, 2015c), it was suggested that PAH's having carbon pentagon-hexagon coupled cation molecules show good coincidence with observed IR. Typical example was coronene based void induced di-cation $C_{23}H_{12}^{2+}$ having two carbon pentagons combined with five hexagons. It was amazing that this single molecule could almost reproduce IR spectrum with astronomically well observed one in a wavelength range from 3 to 15 micrometer. Here, in order to clarify such coincidence, a capable astronomical chemical evolution path from graphene to PAH was studied based on an astrophysical recent understanding on star, gas and dust. Here scientific tool is a computer to analyze model molecules by quantum chemistry based density functional theory (DFT) method. There are two ways to understand chemical evolution in the interstellar space. One is so called bottom-up chemical evolution hypothesis, that is, one carbon bonds with other carbon and/or hydrogen, two carbon molecule grow to three, four and more complex one. Another hypothesis is top-down model, which concept was illustrated in a review titled "Molecular Universe" by Tielens (Tielens 2013). Unfortunately, there are no detailed and quantitative studies until now. Starting material will be large size carbon dust which will be modified by astronomical photons and particles.

Here, I like to take my stand on top-down model especially focusing on nano-carbon. The first step would be relatively large size molecular dust ejected from a star as like diamond, graphite and/or graphene molecules. Those would be chemically modified by high energy particles (mostly proton and electron) and photon. In this study as a typical carbon dust model, we like to choose graphene molecule. For simplicity, small molecule $C_{24}$ with coronene skeleton having seven carbon hexagons was calculated. Step 1 is the nucleation of nano-carbon after supernova. Step 2 is proton $H^+$ sputtering and passivation in gas cloud. Step 3 is photo-ionization both in case of supernova decayed photo emission and new born star illumination. Every energy level of model molecule will be compared based on the DFT. Analyzed energy tells us quantitatively the proton dynamic energy for hydrogenation and void creation. Photon energy diagram will show us how about central star effective temperature and relating star size. Finally, we like to obtain calculated IR spectrum comparing astronomically observed one.

## 2, CALCULATION METHOD

We have to obtain total energy, optimized atom configuration, and infrared vibrational mode frequency and strength depend on a given initial atomic configuration, charge and spin state Sz. Density functional theory (DFT) with unrestricted B3LYP functional (Becke 1993) was applied utilizing Gaussian09 package (Frisch et al. 2009, 1984) employing an atomic orbital 6-31G basis set. The first step calculation is to obtain the self-consistent energy, optimized atomic configuration and spin density. Required convergence on the root mean square density matrix was less than $10^{-8}$ within 128 cycles. Based on such optimized results, harmonic vibrational frequency and strength was calculated. Vibration strength is obtained as molar absorption coefficient ε (km/mol.). Comparing DFT harmonic wavenumber $N_{DFT}$ ($cm^{-1}$) with experimental data, a single scale factor 0.965 was used (Ota 2015b). For the anharmonic correction, a redshift of $15cm^{-1}$ was applied (Ricca et al. 2012).

Corrected wave number N is obtained simply by $N (cm^{-1}) = N_{DFT} (cm^{-1}) \times 0.965 - 15 (cm^{-1})$.
Also, wavelength λ is obtained by $\lambda (\mu m) = 10000/N(cm^{-1})$.

## 3, MODELING CHEMICAL EVOLUTION PATH

### Step 1, Nucleation of carbon dust after supernova

Origin of interstellar dust was assumed to be supernova related phenomenon. Recently, theoretical astrophysical scientist challenged the nucleation mechanism of dust. As shown in left side of Figure 1, the heavy old star with mass larger than ten times of our sun has an onion like internal structure by nuclear fusion mechanism. The most outside shell is hydrogen, next helium, carbon and oxygen, neon and magnesium and finally iron central core. There happens supernova explosion and brings nucleation of supernova remnant. Nozawa et al. (Nozawa 2003, 2006) simulated nucleation mechanism. As illustrated in Figure 1, supernova outmost hydrogen and helium sphere drastically expand to space as adiabatic expansion. There happens super-cooling inside of helium sphere and brings nucleation of carbon dust. In case of 20 times solar mass star, after 300 days of explosion carbon dust will be nucleated and ejected to outer space.

Size of carbon dust was calculated to be in a range of 5nm to 1000nm (Nozawa 2003). Average dust size will be 100nm. In this study, we like to model smaller 1 nm size nano-carbon. The reason is that we could imagine carbon dust to be an aggregate or amorphous composed by nanometer size carbon molecules. Here, we like to focus on graphene molecule, especially on simple molecule coronene skeleton $C_{24}$ having seven carbon hexagons, which molecular structure is illustrated in a down right side of Figure 1.

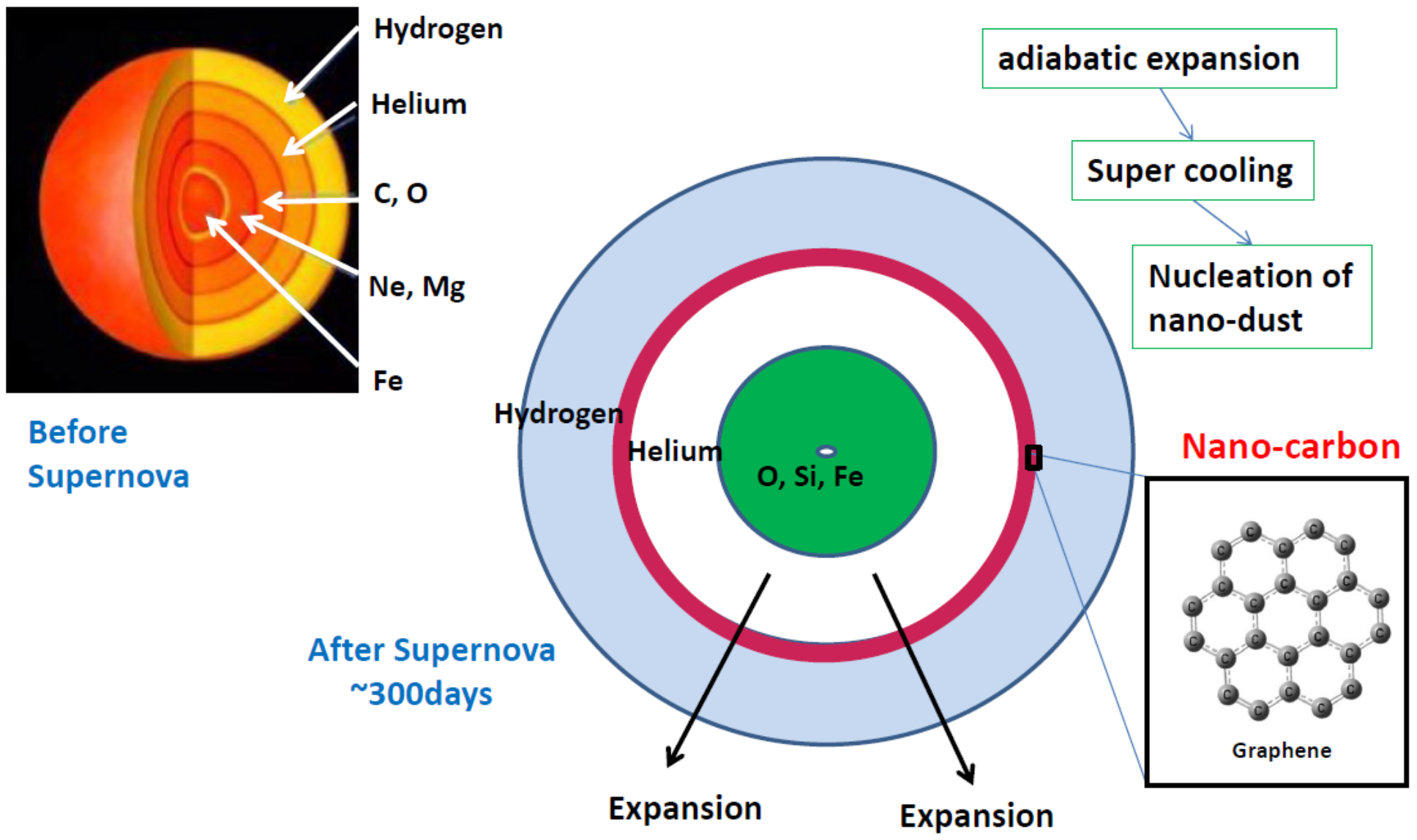

Figure 1, Nucleation of nano-carbon after supernova expansion. During adiabatic expansion, there occurs super cooling in helium sphear and brings carbon nucleation. An typical example of nano-carbon is graphene molecule.

**Step 2, Interstellar proton $H^+$ gas sputtering on nano-carbon**

Ejected nano-carbon from supernova diffuses to surrounding space at a high speed and will collide with interstellar gas, mainly with interstellar proton $H^+$. In a sense of material science, this means sputtering and/or passivation on nano-carbon by proton. In Figure 2, molecular configuration change by proton sputtering was illustrated. Starting is graphene molecule $C_{24}$ as shown in (a). By low speed proton sputtering, there occurs hydrogenation of molecular edge. Fully hydrogenated one is $C_{24}H_{12}$ as noted in (b). Each carbon edge is chemically radical and easily modified by proton $H^+$. Calculated bonding energy for C-H was 4.3eV. This also suggests high kinetic energy proton larger than 4.3eV will bring dehydrogenation of C-H bond. In case of high speed proton sputtering, carbon atom inside of a molecule will be attacked and kicked out. A void will be created inside of a molecule as illustrated in (c) as $C_{23}H_{12}$. Void induced molecule would be transformed to a quantum mechanically stable structure. DFT based calculation resulted a stable one to be shown in (d). Stabilization energy of $C_{23}H_{12}$ was -5.2eV. Molecular structure is a combination of two carbon pentagons combined with five hexagons.

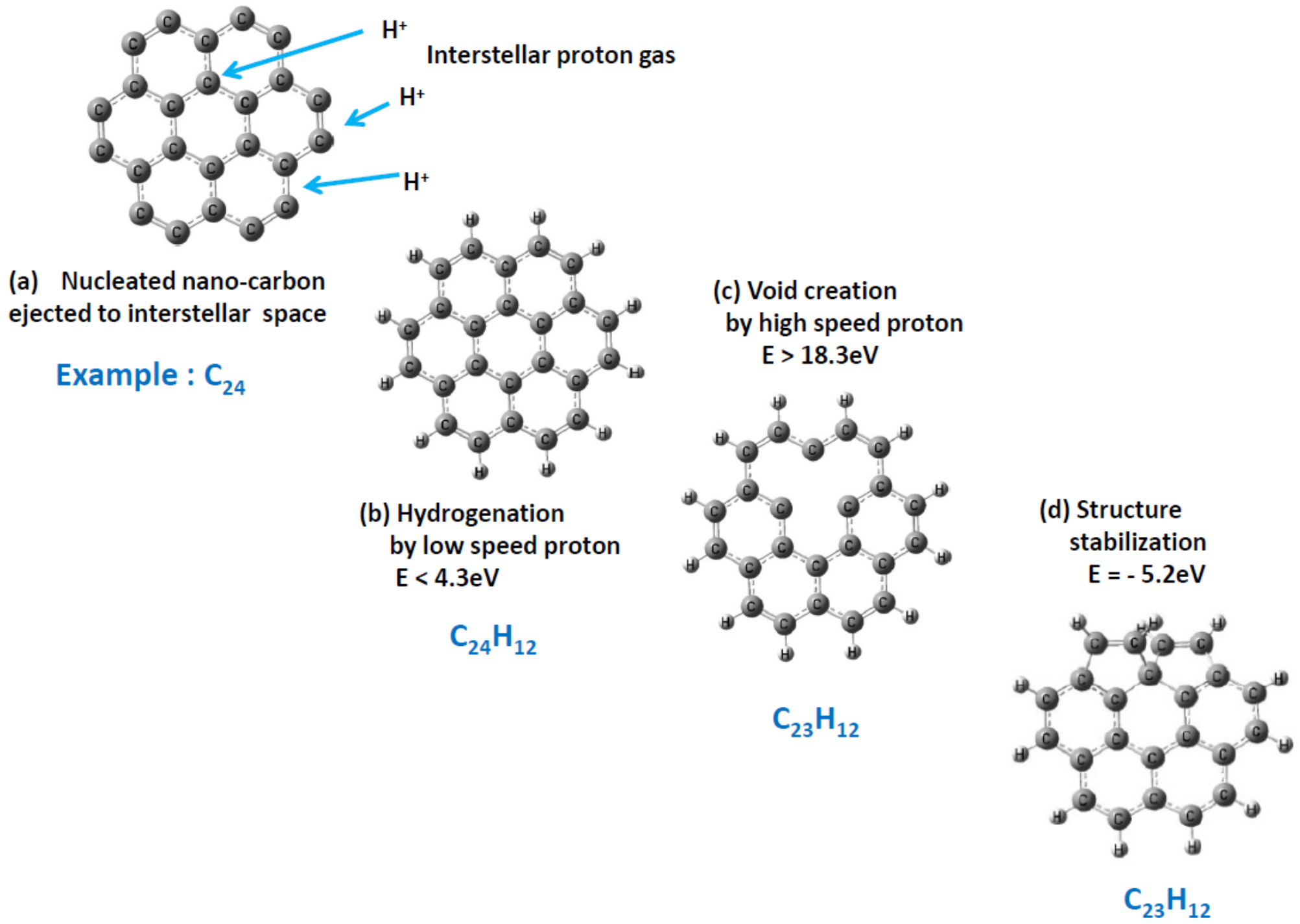

Figure 2, Proton ($H^+$) sputtering on nucleated graphene molecule after supernova in interstellar space

**Step 3, Photoionization**

Photoionization is the essential step for interstellar dust chemical evolution, especially reproducing interstellar PAH. There may be two capable photon sources in a life of star. One is decayed light illumination from heavy star after supernova. In Figure 3 bottom left, we can see central star of supernova SN1987A illuminating remnant gas and dust after 9 year from explosion, which was observed by Hubble space telescope (NASA opened photo). As illustrated in Figure 3, ejected dust in supernova remnant ring would be photo-ionized by suitable energy photon. In case of $C_{23}H_{12}$ molecule, electrons in molecule will be stripped away one by one by high energy photon. Another photon radiation source capability is new born star. As shown in Figure 4, we can see famous Rosette Nebula (NGC2237) opened by NASA. Central new born star is illuminating surrounding dust cloud. In such a case too, electrons of dust molecule will be stripped away, that is, molecule $C_{23}H_{12}$ may change to cation $C_{23}H_{12}^{n+}$ (n=1, 2, 3…).

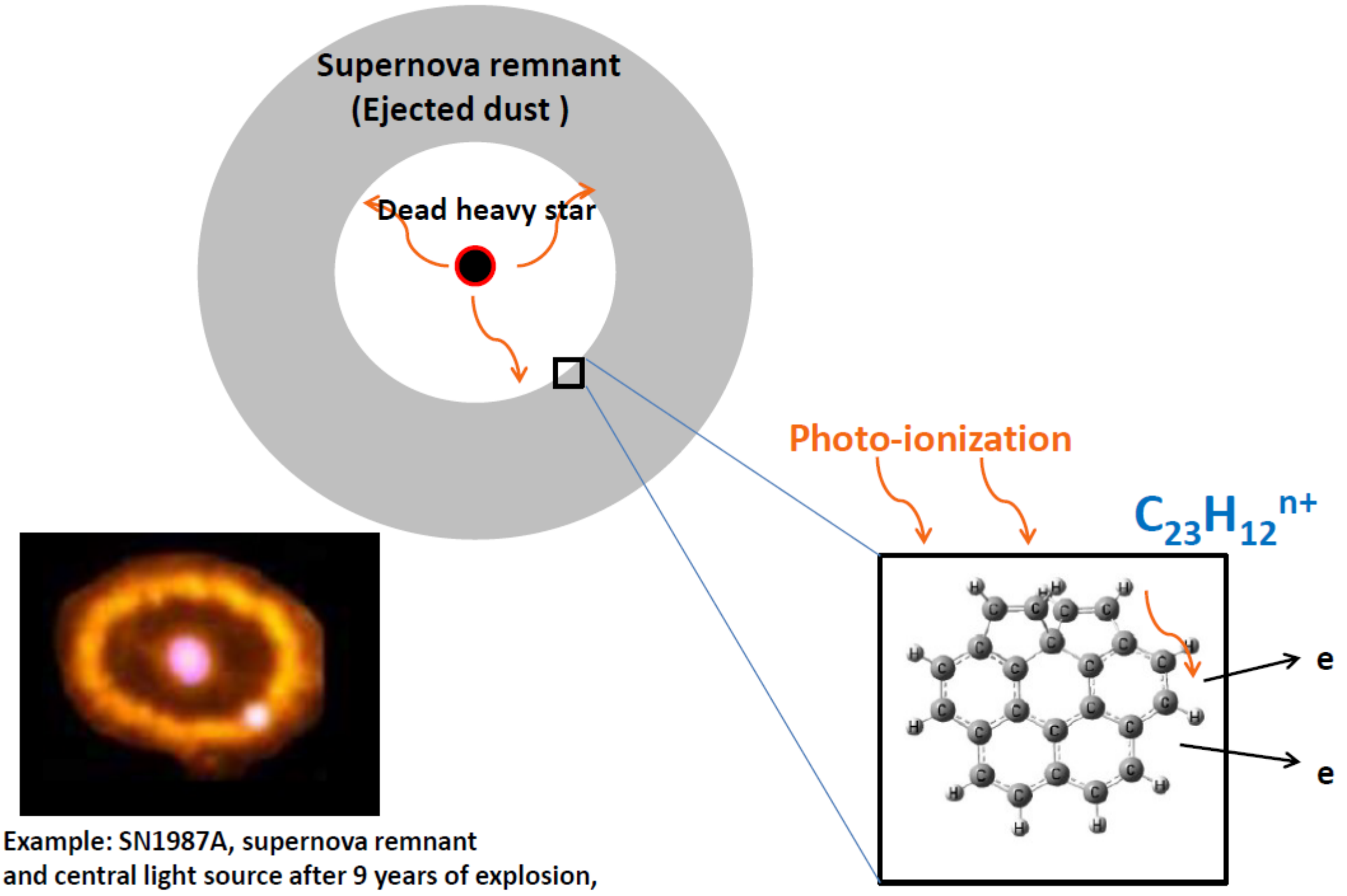

Example: SN1987A, supernova remnant
and central light source after 9 years of explosion,
Hubble Space Telescope (NASA)

Figure 3, Photoionization after supernova by decayed light from the central heavy star.

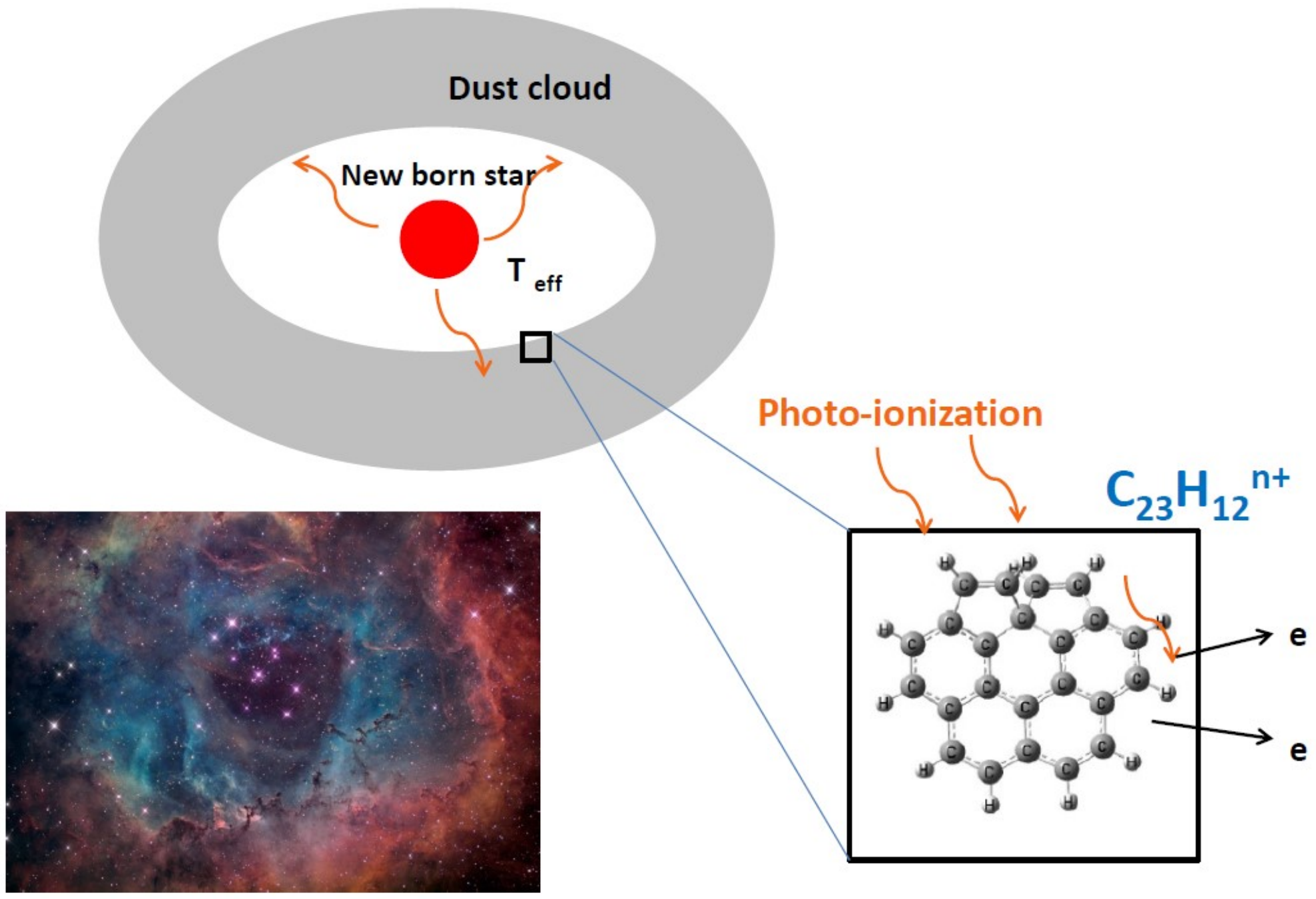

Example : The Rosette Nebula (NGC2237) , credit and edited by NASA

Figure 4, Photoionization by new born star radiation

## 4, ENERGY LEVEL DIAGRAM AND CENTRAL STAR EFFECTIVE TEMPERATURE

According to a top down chemical evolution hypothesis, energy level change of model graphene was calculated and illustrated in Figure 5. Graphene molecule $C_{24}$ was fully hydrogenated to $C_{24}H_{12}$ (coronene). Void creation energy to $C_{23}H_{12}$ was calculated to be 18.31eV. Quantum chemically stabilized energy for molecular configuration change was -5.24eV. Molecular energy of stabilized one with neutral charge was defined as $E_0$=0eV. Starting from neutral one, cationic molecules configuration and energy were calculated by spin dependent DFT method. Mono cation $C_{23}H_{12}^{+}$ was obtained by 6.50eV higher energy as $E_1$, which configuration is similar with neutral one. In a previous study (Ota 2014), mono-cation configuration was different with this study. It was recalculated that the most stable (lowest energy) molecular configuration is similar with this study's neutral one. From mono- to di-cation $C_{23}H_{12}^{2}$, energy difference ($E_2$-$E_1$) was 10.66eV. Similarly, ($E_2$-$E_3$) was 14.52eV, ($E_3$-$E_4$) 18.59eV. We could obtain a stable configuration even in high cationic state $C_{23}H_{12}^{4+}$ having structure of two carbon pentagons combined with five hexagons.

If the central star as a photon source had the black body radiation nature, we could estimate the effective temperature based on the Plank's rule. Figure 6 shows radiation curve with effective temperature $T_{eff}$ for each 18000, 20000, 22000, and 24000K. Horizontal axis shows photon energy by eV unit. We could directly compare with photo-ionization energy ($E_1$-$E_0$), ($E_2$-$E_1$), ($E_3$-$E_2$), and ($E_4$-$E_3$) noted by blue arrows. Also, direct excitation from $E_0$ to $E_2$ was marked by orange broken arrow ($E_2$-$E_0$). It should be noted that such energy excitation could be satisfied by photon radiation from a source following Plank's curve with Teff =18000 ~24000K. If the central star was the main sequence star, such effective temperature suggests that star mass would be 4 to 7 times heavier than our sun.

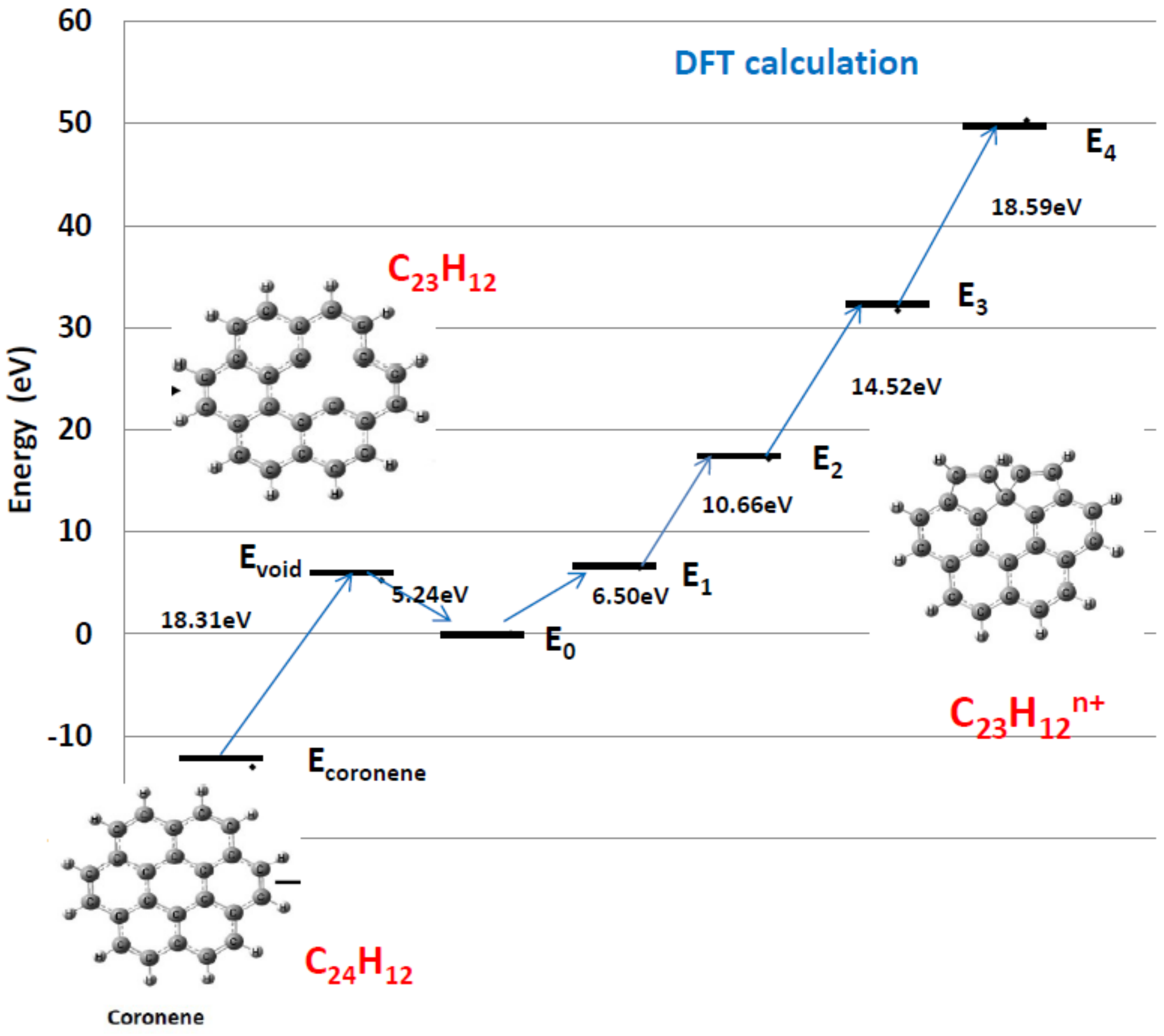

Figure 5, Energy level diagram of $C_{24}H_{12}$ and $C_{23}H_{12}^{n+}$ (n=0, 1, 2, 3, and 4).

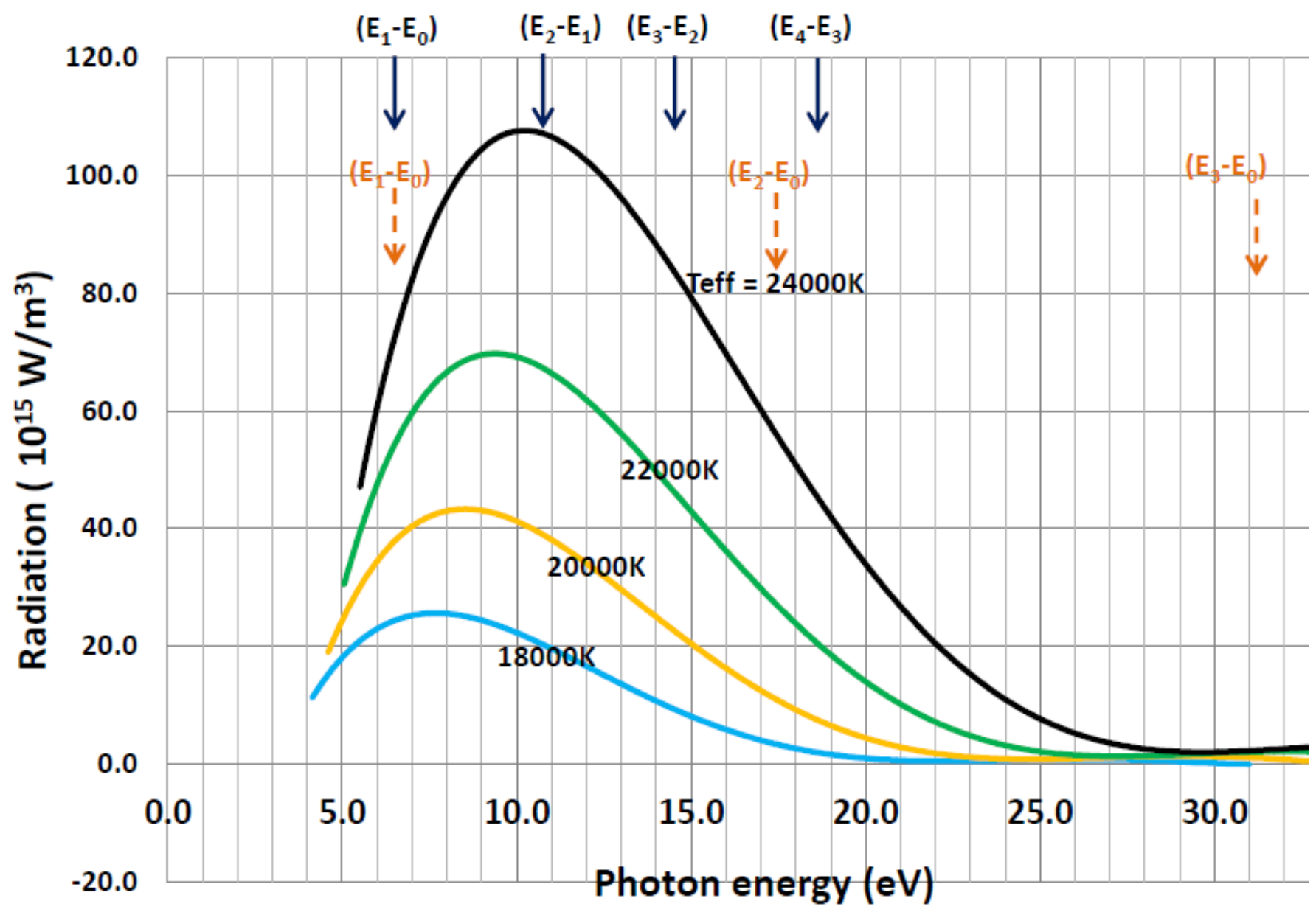

Figure 6, Photon energy profile by black body radiation Plank's rule with effective temperature $T_{eff}$=18000~24000K. Photoionization energy for $C_{23}H_{12}{}^{n+}$ molecules is noted by arrow. If the central star were main sequence star, mass would be 4~7 times heavier than our sun.

## 5, INFRARED SPECTRUM

Infrared spectrum was calculated for each ionized molecules $C_{23}H_{12}{}^{n+}$.

(1) Neutral $C_{23}H_{12}$

Molecular configuration was illustrated in right side of Figure 7 as a front view and a side view. Angle between two carbon pentagons θ (C1-C2-C3) was 94.7 degree. Permanent electric dipole moment D was 0.65 Debye, which positioned at the molecular center in the upward direction. Molecular vibration was calculated and analyzed as infrared spectrum as shown in top left. Major wavelength was 3.3, 11.6, 12.0, and 17.6 micrometer.

(2) Mono cation $C_{23}H_{12}{}^{+}$

As shown in Figure 8 (b), fundamental configuration is almost similar with neutral one, but somewhat increase in angle θ as 95.0 degree. Dipole moment D was 0.91 Debye. Calculated IR was close to astronomically observed one as compared in a following table. However, there remains somewhat discrepancy within +/- 0.4 micrometer between observation and calculation.

| Calculation (micrometer) | 3.2 | 6.4 | 7.2 | 7.5 | 8.3 | 8.6 | 9.0 | 11.6 |
|---|---|---|---|---|---|---|---|---|
| Observation | 3.3 | 6.2 | 7.6 | 7.8 | none | 8.6 | none | 11.3 |

(3) Di-cation $C_{23}H_{12}{}^{2+}$

In Figure 8(c), angle θ was increased to 95.4 degree and dipole moment was also increased to D=1.20 Debye. It should be noted that calculated IR became more close to observed one as summarized in a following table.

| Calculation (micrometer) | 3.2 | 6.4 | 7.2 | 7.6 | 7.8 | 8.2 | 8.6 | 11.3 | 12.2 | 12.8 | 14.2 |
|---|---|---|---|---|---|---|---|---|---|---|---|
| Observation | 3.3 | 6.2 | none | 7.6 | 7.8 | none | 8.6 | 11.3 | 12.0 | 12.7 | 14.2 |

This was the same result with previously reported one (Ota 2014).

(4) Tri-cation $C_{23}H_{12}{}^{3+}$

Result of tri-cation was shown in Figure 8(d). Angle θ was 99.0 degree. Dipole moment was small D=0.22 Debye.

Major wavelength of calculated IR was 3.2, 6.7, 7.7, 8.6, 9.6, 11.3, and 12.9 micrometer, which suggested large discrepancy in 6 micron band and no observation in 9 micron band.

(5) Quadri-cation $C_{23}H_{12}^{4+}$

Direction of dipole moment was opposite as shown in Figure 8(e) with D= -0.60 Debye. Major wavelength was 3.2, 6.7, 7.7, 11.1, and 13.1 micrometer, which shows large discrepancy in 6 micron band with observed one.

It could be summarized that such molecular family from mono- to quadri- cation $C_{23}H_{12}^{n+}$ was a firstly suggested candidate to reproduce ubiquitous IR of interstellar PAH.

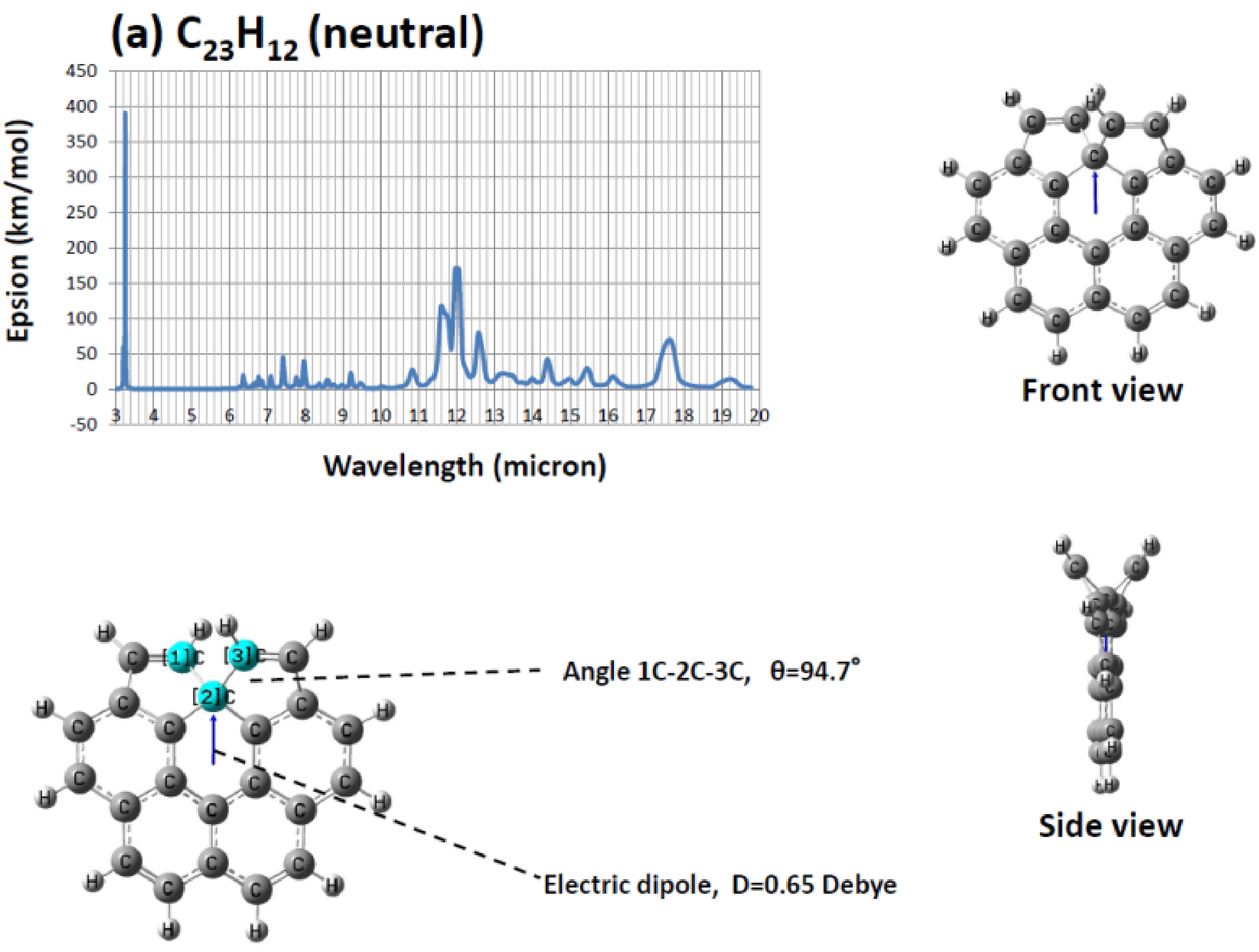

Figure 7,Calculated molecular structure, electric dipole moment D and IR of neutral $C_{23}H_{12}$ molecule

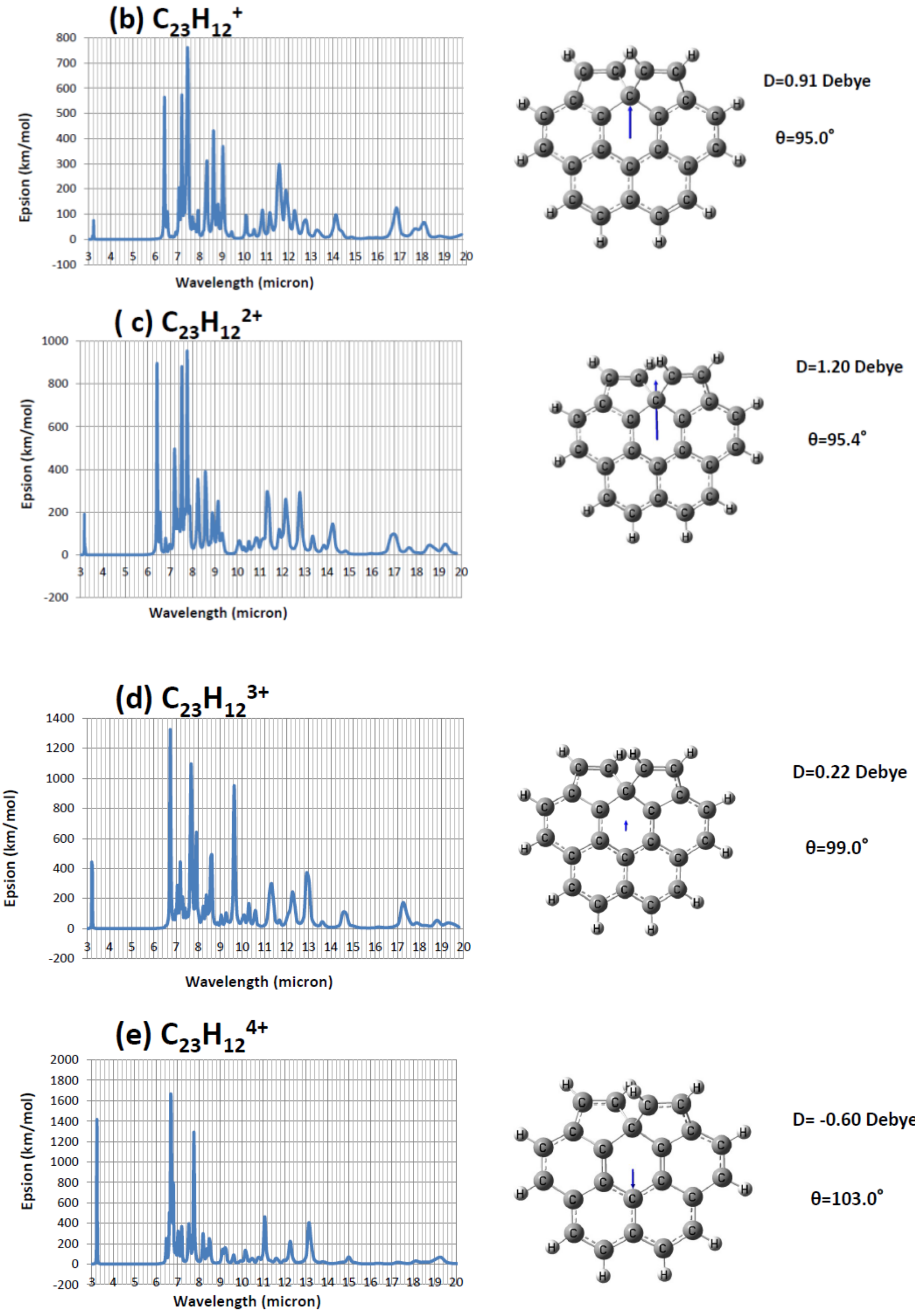

Figure 8, Calculated molecular structure, electric dipole moment D and IR of cationic molecules $C_{23}H_{12}^{n+}$ (n=1, 2, 3, and 4).

## 6, OBSERVED AND CALCULATED EMISSION INFRARED SPECTRUM

Among cation family, the most promising one was di-cation $C_{23}H_{12}^{2+}$. Observed IR was emission spectrum. Emission infrared spectrum should be calculated and compared with them. Dr. Christiaan Boersma of NASA Ames Research very kindly calculated emission spectrum based on my DFT based fundamental vibrational mode analysis. Figure 9 shows his calculation result in red. For comparison, observed emission IR's from typical four astronomical sources are illustrated on top (Boersma et al. 2009). Dust cloud of HD44179 is 2300 light years away from the earth, which is considered to be white dwarf, oldest star in star life stage. Whereas, NGC7027 is young star, 3000 light years away. Both clouds around young and old star shows similar IR tendency, which may suggest two capabilities in photoionization as noted in this paper. Comparing with observed IR and calculated emission one, we can find best agreement in wavelength of 3.2, 7.6, 7.8, 8.6, 11.2, 12.0, 12.7, and 14.2 micrometer. However observed 6.2 micrometer band was calculated to be 6.4 micrometer. Also, calculated emission strength of 11.2 micrometer band is relatively half of observed one. Those minor differences may suggest that in order to obtain complete coincidence we should consider some nitrogen, oxygen and/or aliphatic component inclusion and modification (Ota 2015b ). Calculated extra band of 8.2 and 9.0 micrometer were not observed as PAH band. It remains future advanced study.

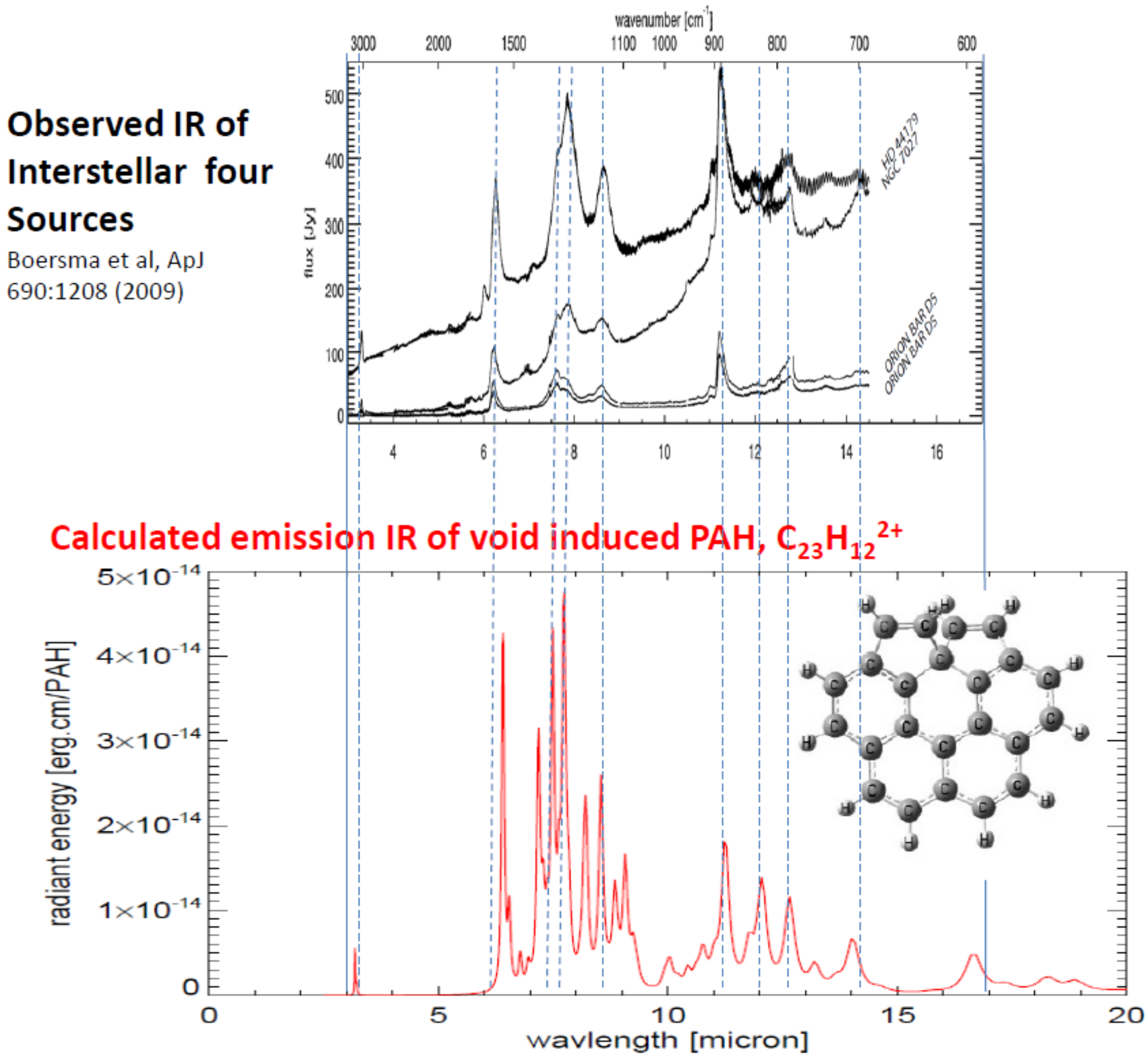

Figure 9, Observed interstellar spectrum from four astronomical sources (top illustrated curves) and calculated emission IR spectrum kindly provided by Dr. Chrstiaan Boersma (down in red)

## 6, CONCLUSION

Interstellar ubiquitous infrared spectrum (IR) due to polycyclic aromatic hydrocarbon (PAH) was observed in many astronomical dust clouds. In a previous quantum chemistry study, it was suggested that PAH's having carbon pentagon-hexagon coupled molecules show good coincidence with observed IR. In order to clarify such coincidence, a capable astronomical chemical evolution path from graphene to PAH was studied based on first principles calculation.
(1) Step 1 is a nucleation of nano-carbon after super-nova expansion by the super-cooling at expanding helium sphere. Graphene molecule $C_{24}$ with coronene skeleton is a typical model.
(2) Step 2 is a proton sputtering and passivation on ejected graphene molecule in a proton gas cloud. Slow speed proton makes hydrogenation. Graphene molecule $C_{24}$ was transformed to PAH $C_{24}H_{12}$. Higher speed proton could make a void in a molecule as like $C_{23}H_{12}$, which molecular structure was a combination of two carbon pentagons combined with five hexagons.
(3) Step 3 is photo-ionization of those molecules by high energy photon. Model molecule $C_{23}H_{12}$ became $C_{23}H_{12}^{+}$, $C_{23}H_{12}^{2+}$ and so on. Typical energy difference between such cation was 6.5 and 10.8 eV. If the light source has a nature of black-body photo-emission, effective temperature will be 18000K ~ 24000K, which suggested that central light source star may have 4 to 7 times heavier than our sun. Finally, theoretical IR spectrum was obtained for such cation. Very nice coincidence with observed IR was obtained. Especially in case of $C_{23}H_{12}^{2+}$, calculated emission spectrum revealed entire coincidence with many observed IR from 3 to 15 micrometer.

## ACKNOWLEDGEMENT

I would like to say great thanks to Dr. Christiaan Boersma, NASA Ames Research Center to provide me an emission IR calculation illustrated in Figure 9 based on my DFT based fundamental vibrational mode analysis. It's a great present for me on Dec. 25, 2014. Also, I would like to appreciate him to permit me to refer an important observed IR figure (Boersma et al. 2009).

Submitted to arXiv.org on March , 2017 by Norio Ota